\documentclass[acmlarge]{acmart}

\AtBeginDocument{%
  }

\begin{document}

\title{Exploring Image Quality Assessment from a New Perspective: Pupil Size}

\author{Yixuan Gao}
\email{gaoyixuan@sjtu.edu.cn}
\author{Xiongkuo Min}
\authornote{Corresponding authors.}
\email{minxiongkuo@sjtu.edu.cn}
\author{Guangtao Zhai}
\authornotemark[1]
\email{zhaiguangtao@sjtu.edu.cn}
\affiliation{%
  \institution{Institute of Image Communication and Network Engineering, Shanghai Jiao Tong University}
  \city{Shanghai}
  \country{China}
}

\renewcommand{\shortauthors}{Gao et al.}
\begin{abstract}
  This paper explores how the image quality assessment (IQA) task affects the cognitive processes of people from the perspective of pupil size and studies the relationship between pupil size and image quality. Specifically, we first invited subjects to participate in a subjective experiment, which includes two tasks: free observation and IQA. In the free observation task, subjects did not need to perform any action, and they only needed to observe images as they usually do with an album. In the IQA task, subjects were required to score images according to their overall impression of image quality. Then, by analyzing the difference in pupil size between the two tasks, we find that people may activate the visual attention mechanism when evaluating image quality.
Meanwhile, we also find that the change in pupil size is closely related to image quality in the IQA task.
For future research on IQA, this research can not only provide a theoretical basis for the objective IQA method and promote the development of more effective objective IQA methods, but also provide a new subjective IQA method for collecting the authentic subjective impression of image quality.
\end{abstract}




\maketitle

\section{Introduction}
Recently, image quality assessment (IQA) has become a hot research field \cite{wang2004image,zhang2011fsim,mittal2012no,2018Blind,2012A,gao2022imageTCSVT}. Specifically, IQA can be divided into subjective IQA \cite{LIVE,virtanen2014cid2013,hosu2020koniq} and objective IQA \cite{radun2008content,gao2017feature,min2017blind,zhai2020perceptual,gao2022image}. Subjective IQA measures image quality by integrating subjective opinion scores given by subjects. 
Objective IQA computes image quality through some computational models, which generally model human vision. For example, the contrast sensitivity function, just noticeable difference model, and visual attention mechanism are frequently used to model the human visual system (HVS) to calculate image quality \cite{fang2019image,gao2009image,liu2011visual,cao2023attention}.
 The successful applications of these HVS-based objective IQA methods are benefited from sufficient research on human visual perception.

Vision is an important way for people to obtain and perceive image information. HVS can obtain visual information by controlling the eye movement or the constriction and dilation of the pupil, and transmit the obtained information to the brain for further analysis. 
As the gateway through which people observe the world, the pupil plays an important role in human visual perception. The pupil is in the center of the iris.
Changes in pupil size, that is, pupil constriction and dilation, can control the amount of light emitted to the retina, which reflect the basic physiological reactions of human vision caused by changes in ambient brightness \cite{woodhouse1975role}. In addition, changes in pupil size can also reflect the task-evoked cognitive processes of people. For example,
Kahneman \emph{et al}. \cite{kahneman1966pupil} analyzed the task-evoked pupil size in short-term memory tasks. They found that both the difficulty and number of tasks affect pupil size.
Rieger \emph{et al}. \cite{rieger2012eyes} proposed that the pupil dilates in response to attractive social partners. Some studies on the relationship between emotional stimuli and pupil size showed that the dilated pupil indicates a positive stimulus and the constricted pupil indicates a negative stimulus \cite{hess1960pupil,mudd1990eye}.
Authors in \cite{sabatino2018task} explored the differences in pupil responses to local and global information. They found that pupils constrict significantly when people focus on local information rather than global information. 
Generally speaking, changes in pupil size are different for different tasks. Therefore, this paper first explores how the IQA task affects the cognitive processes of people from
the perspective of pupil size.

The performance of objective IQA methods is usually measured by the correlation between the calculated image quality and the image quality obtained from subjective experiments. In other words, both subjective IQA methods and objective IQA methods are sensitive to the subjective opinion scores collected from subjects. 
Recently, researchers have found that the subjective opinion scores collected by the five-level rating scale proposed by ITU-R \cite{1} may have some errors in reflecting the authentic human visual perception brought by images \cite{laparra2009emg}.
For example, some subjects may be dishonest when evaluating image quality.
In order to solve this problem, 
researchers have collected people's authentic impressions of image quality by studying human physiological signals. Al-Juboori \emph{et al}. \cite{al2017investigation} investigated the relationship between changes in the electroencephalogram and subjective quality scores for high dynamic range images.
Gutu \emph{et al}. \cite{gutu2014analysis} studied the relationship between facial electromyography and image quality.
Although these methods are effective, it is inconvenient to carry out subjective experiments because of the heavy apparatuses. In contrast, pupil size can be measured with a portable eye tracker. In addition, some studies have proved that the pupil responds to both the low-level (color and structure in an image) and high-level (content semantics) processing in the HVS \cite{1972Pupillometrics,pappusetty2017using}. 
Therefore, this paper also studies whether different image quality levels can be measured by changes in pupil size when people observe images.


Our work has following contributions. First, this paper explores how the IQA task affects the cognitive processes of people from the perspective of pupil size, which can provide a solid theoretical basis for the objective IQA method and promote the development of more effective objective IQA methods.
Second, the relationship between the change in pupil size and image quality is investigated, 
which can be used to 
guide the subjective IQA method to collect the authentic subjective impression of image quality using pupil size.

The remaining framework of the paper is organized as follows. In Section \ref{subjective Experiment}, we introduce the subjective experiment in detail.
Section \ref{Analysis of Pupil Data} analyzes the obtained pupil data.
In Section \ref{S4}, we explore the relationship between image quality and pupil size.
Finally, a summary of the paper is given in Section \ref{S5}.

\begin{figure*}[t]
\vspace{-40pt}
\begin{minipage}[b]{1.0\linewidth}
  \centering
  \vspace{1.5cm}
  \centerline{\includegraphics[scale=0.8, trim=20 135 470 120, clip]{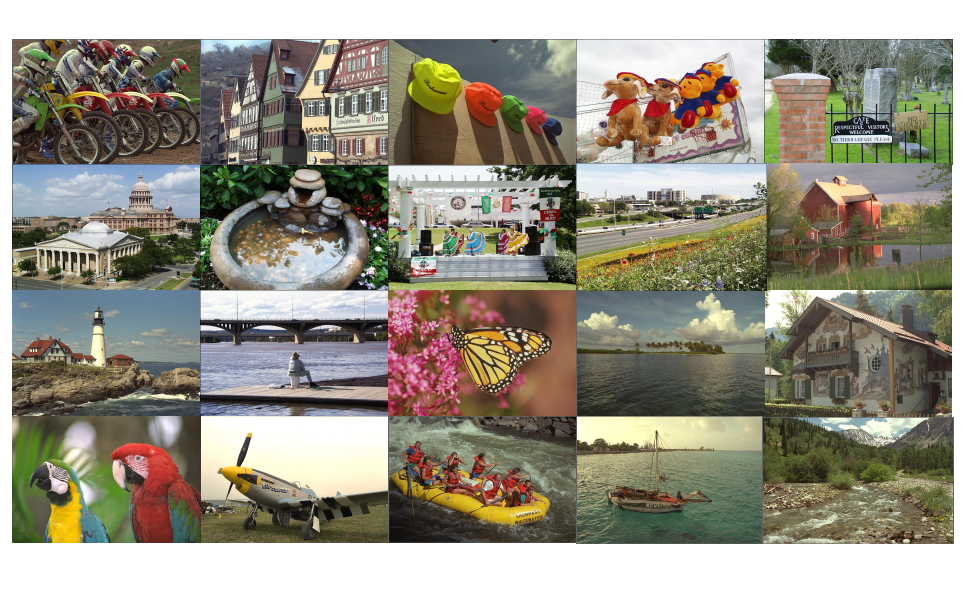}}
\end{minipage}
\vspace{-18pt}
\caption{20 reference images selected from the LIVE database.
}
\label{ref}
\end{figure*}

\begin{figure}[t]
\begin{minipage}[b]{1.0\linewidth}
\vspace{-40pt}
  \centering
  \vspace{1.5cm}
  \centerline{\includegraphics[scale=0.55, trim=80 250 80 280, clip]{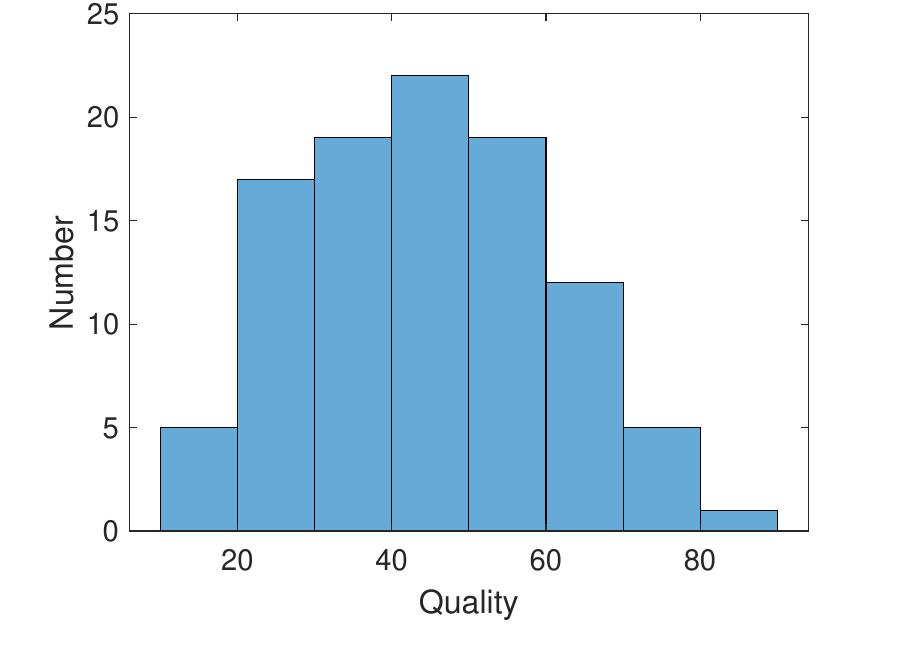}}
\end{minipage}
\vspace{-18pt}
\caption{Distribution of quality of all selected distorted images provided by the LIVE database.}
\label{LIVE_quality.}
\end{figure}

\section{Subjective Experiment}
\label{subjective Experiment}
The subjective experiment includes two tasks: free observation and IQA. 
The details of the experiment, including stimulus materials, subjects, apparatus, and experimental procedures, are described as follows.
\subsection{Stimulus}
The stimulus materials used in the subjective experiment were 100 distorted images selected from the LIVE database \cite{LIVE}. Specifically, we first selected 20 reference images from the LIVE database. Fig. \ref{ref} shows the 20 reference images.
Then, we selected five distorted images with different quality levels for each reference image. Thus, we have a total of 100 distorted images.
These distorted images are affected by five types of distortions, including JPEG2000 (JP2K), JPEG, white noise (WN), Gaussian blur (Gblur), and fast-fading Rayleigh (FF). There are 20 distorted images for each distortion type.
Fig. \ref{LIVE_quality.} shows the distribution of quality of all selected distorted images provided by the LIVE database.
\subsection{Subjects}
There were a total of 17 subjects in the subjective experiment. They were all students, about 18 to 30 years old. All subjects had normal (corrected) vision and no eye disease. This is considered sufficient to ensure that they can observe the differences between images and that we can track the normal changes in their pupil size. All subjects were naive about the purpose of the subjective experiment.
\subsection{Apparatus}
A $1920\times 1080$ 23.8-inch monitor was used to display all images, which has a peak luminance of 250 $cd/m^{2}$ and a refresh rate of 60 Hz. Note that the subjective experiment was conducted in a laboratory with normal indoor lighting.
The viewing distance between the monitor and the subjects was about 70 centimeters.
We used the Tobii Pro X3-120 eye tracker with a maximum sampling rate of 120 Hz to record pupil size. The eye tracker is specially designed for the detailed study of natural behavior, whose large head movement box allows the subject's head to move within a certain range during the recording process while the maintaining accuracy and precision of the recording. Therefore, the heads of the subjects did not need to be fixed during the experiment.
The experimental program was created using Tobii Pro Lab, and the standard five-point calibration procedure was used in Tobii Pro Lab.

\begin{figure*}[t]
\begin{minipage}[(b) ]{0.45\linewidth}
  \centering
  \centerline{\includegraphics[scale=0.45, trim=20 100 500 60, clip]{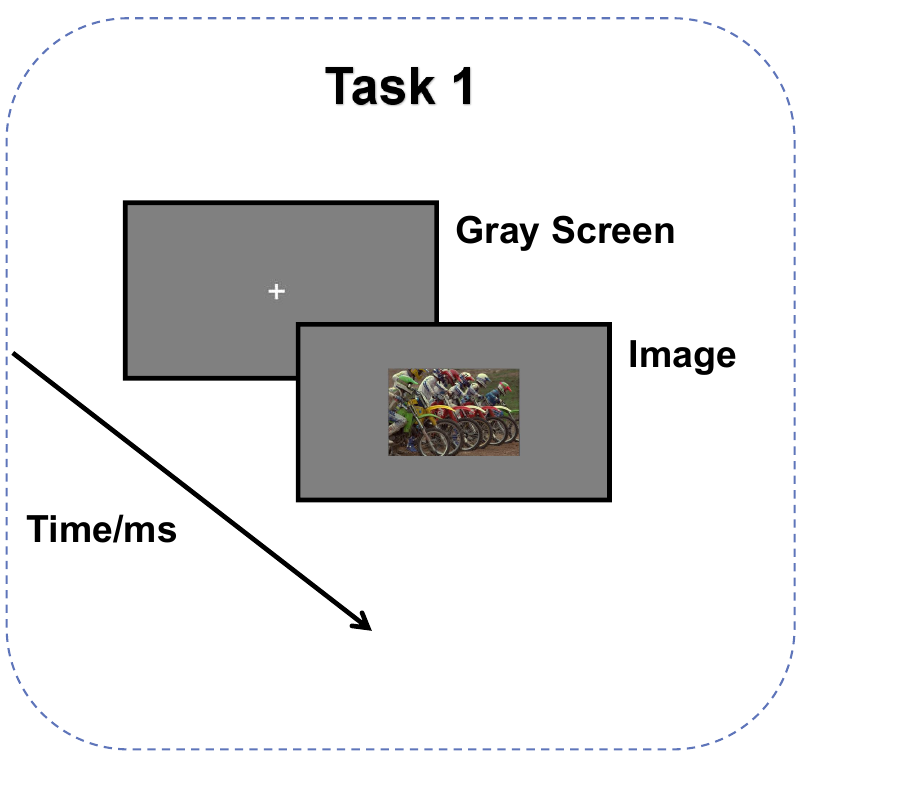}}
  \centerline{(a) Task 1: Free observation.}\medskip
  \end{minipage}
\begin{minipage}[(b) ]{0.45\linewidth}
  \centering
  \centerline{\includegraphics[scale=0.45, trim=20 100 490 70, clip]{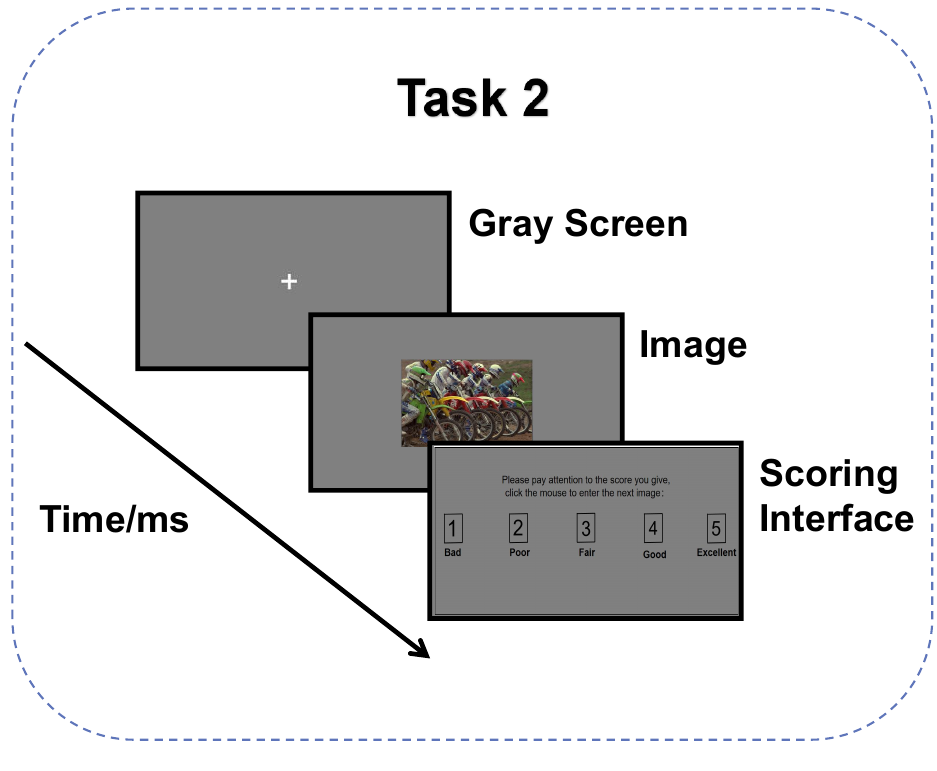}}
  \centerline{(b) Task 2: IQA.}\medskip
\end{minipage}
\caption{Experimental procedures. (a) shows the display procedure of stimulus materials in Task 1. (b) shows the display procedure of stimulus materials in Task 2. 
}
\vspace{-5pt}
  \label{sequence}
\end{figure*}
\subsection{Experimental Procedures}
\label{sec:2}
The subjective experiment included two independent tasks. The first task was free observation. In this task, all subjects did not need to perform any action, and they only needed to observe images freely. The second task was IQA. In this task, subjects needed to give their subjective opinion scores of image quality after observing images. 
The interval between the two tasks was long enough to avoid fatigue which may affect the pupil size of subjects.
The experimental procedures for the two tasks are shown in Fig. \ref{sequence} and described in detail below.

\subsubsection{Task 1: Free observation}
Before the task, subjects were told that they would participate in an experiment on image observation, and their pupil size would be recorded by an eye tracker. They were required that they only needed to observe images as if they were observing an album at ordinary times throughout the task. 
Then, there was a short training to show subjects five images so that they could become familiar with the experimental process. After the subjects were ready, we performed the calibration process of the eye tracker on them, and then began to show the subjects the stimulus.
The experimental procedure is shown in Fig. \ref{sequence}(a), including a gray screen and an image.
Specifically,
before displaying each image, subjects would be shown a 3s gray screen. Then, in the center of the gray screen, we randomly displayed an image for the subjects. Subjects were free to observe the images for no time limit. After observing an image, subjects could click anywhere on the screen with the mouse to switch to the next stimulus.
At this time, the system would bring subjects to the gray screen again. After 3s, the system randomly selected another image to display. We repeated these steps until the subjects had observed all 100 images.
\subsubsection{Task 2: IQA}
Before the formal experiment began, subjects were told that each image should be scored according to the overall impression of image quality. The correlation between the scoring standard and the ITU-R quality standard \cite{1} is as follows: a score of 1 represents that the image quality is `Bad'; 
a score of 2 represents that the image quality is `Poor'; 
a score of 3 represents that the image quality is `Fair'; 
a score of 4 represents that the image quality is `Good'; 
a score of 5 represents that the image quality is `Excellent'.

We also presented some images to the subjects so that they could familiarize themselves with the approximate range of image quality that they could observe during the experiment. At the same time, we introduced the use of the scoring interface to the subjects.
After the subjects were ready, we performed the calibration process of the eye tracker on them, and then began to show the subjects the stimulus. The IQA task used the single-stimulus assessment method proposed by ITU-R \cite{1}. The experimental procedure is shown in Fig. \ref{sequence}(b), including a gray screen, an image, and a scoring interface. Just like Task 1, before showing each image, subjects would be shown a 3s gray screen. Then, in the center of the gray screen, we randomly displayed an image for the subjects. Different from Task 1, in the IQA task, subjects were asked to carefully observe the displayed image so that they could give their subjective opinion scores on image quality. Subjects could observe the images without time constraints. After observing an image, subjects could click anywhere on the screen with the mouse to switch to enter the scoring interface, where they could see five scores (1 to 5), representing the image quality from bad to excellent. Here, subjects were required to give the quality score of the image immediately, and the system would bring subjects to the gray screen again after clicking the mouse. After 3s, the system randomly selected another image to display. We repeated these steps until subjects had given quality scores for all 100 images.
\subsection{Pupil Data Processing}
We adopt a simple method to process pupil data for both tasks. First, since the eye tracker records the left pupil size and the right pupil size respectively, we take the average of the left pupil size and the right pupil size at the same time as the pupil size at the moment. Second, the missing pupil data are linearly interpolated, and the Hampel filter \cite{pearson2016generalized} is used to correct the abnormal values for each subject.
Two subjects are excluded because they moved excessively during the experiment, which led to the pupil data collected by the eye tracker being less than 75$\%$. Third, to facilitate the comparison between subjects, the entire pupil data of each subject are Z-scored for each task. 
Finally, to analyze the relative change in the pupil size of a subject when observing an image, we subtract the pupil size of the first millisecond from the pupil size at each moment during the observation time of this image to obtain the relative pupil size. 
For convenience, the following pupil size represents the average relative pupil size across all subjects at the same observation time.

\begin{figure}[t]
\begin{minipage}[b]{1.0\linewidth}
\vspace{-40pt}
  \centering
  \vspace{1.5cm}
  \centerline{\includegraphics[scale=0.5, trim=80 250 80 280, clip]{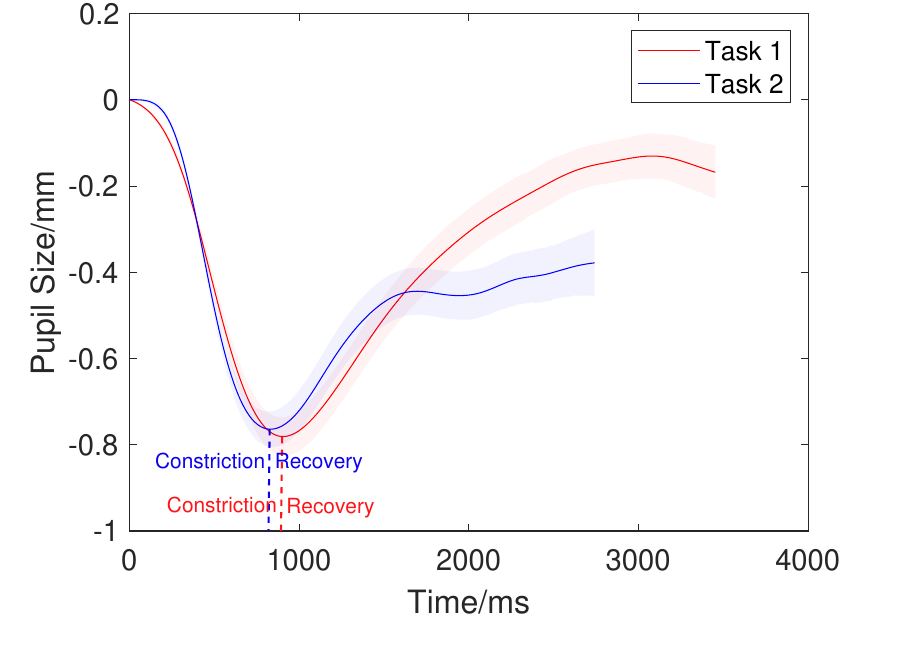}}
\end{minipage}
\vspace{-20pt}
\caption{Changes in pupil size with observation time in Task 1 and Task 2. Light colored areas represent standard errors.
Note that curves are Gaussian smoothed. The dotted line divides the change in pupil size into two phases: constriction and recovery.
}
\vspace{-5pt}
\label{Pupil_Size_all}
\end{figure}

\section{Analysis of Pupil Data}
\vspace{-4pt}
\label{Analysis of Pupil Data}

The collected pupil data shows some interesting differences in the cognitive processes of subjects between Task 1 and Task 2.

First, for each task, we calculate the average observation time across subjects when observing an image. 
Specifically, we consider the pupil size of each subject during the observation of each image as a vector.
Thus, we assume that the pupil size vector for the $i$-th subject during the observation of the $j$-th image can be described as $\textbf{S}_{i}^{j}$, where $i=1,2,\dots,S,$ $j=1,2,\dots,N,$ $S$ is the number of subjects, and $N$ is the number of images. We suppose that the time spent by the $i$-th subject during the observation of the $j$-th image is $T_{i}^{j}$, where $i=1,2,\dots,S$ and $j=1,2,\dots,N.$ The unit of $T_{i}^{j}$ is  millisecond, so $T_{i}^{j}$ has the same length 
as $\textbf{S}_{i}^{j}$.
Thus, we can calculate the average time of observing an image across all subjects:
\begin{eqnarray}\label{QuantileLoss}
AverTime=\sum_{j=1}^{N}\sum_{i=1}^{S} T_{i}^{j}.
\end{eqnarray}

The results show that the average time for a subject to freely observe an image is about 3.452s, while the average time for a subject to observe an image when evaluating image quality is about 2.740s. This observation is consistent with the research result in \cite{alers2011task}, that is, people prefer to spend more time understanding and appreciating images when they are observing freely.

Next, we process the pupil size vectors.
We calculate the pupil size vector when a subject observes an image in the average time:
if $T_{i}^{j}>AverTime$, the pupil data in $\textbf{S}_{i}^{j}$ that exceed the average time are discarded;
if $T_{i}^{j}<AverTime$, the missing pupil data in $\textbf{S}_{i}^{j}$ from $T_{i}^{j}$ to the average time are noted as “$NAN$”. We set the processed pupil size vector as $ \mathcal{S}_{i}^{j}$. Note that the length of each $\mathcal{S}_{i}^{j}~(i=1,2,\dots,S,~j=1,2,\dots,N)$ is equal to $AverTime$.

In total, we have 1700 pupil size vectors:
\begin{eqnarray}\label{QuantileLoss}
\mathcal{S}_{i}^{j}=[S_{i}^{j}(1),S_{i}^{j}(2),\dots,S_{i}^{j}(AverTime)], ~~i=1,2,\dots,S,~j=1,2,\dots,N,
\end{eqnarray}
where $S_{i}^{j}(k)$ represents the pupil size of the $i$-th subject at the $k$-th millisecond when observing the $j$-th image.
Then, we calculate the relative pupil size of a subject when observing an image:
\begin{eqnarray}\label{QuantileLoss}
\mathcal{S}_{i}^{j}=[0,S_{i}^{j}(2)-S_{i}^{j}(1),\dots,S_{i}^{j}(AverTime)-S_{i}^{j}(1)], 
\end{eqnarray}
where $i=1,2,\dots,S$ and $j=1,2,\dots,N.$

Finally, we can obtain the average relative pupil size vector $Aver\mathcal{S}$ across all subjects when observing an image:
\begin{eqnarray}\label{QuantileLoss}
Aver\mathcal{S}=\left[0,\frac{1}{S\cdot N}\sum_{i=1}^{S}\sum_{j=1}^{N}[S_{i}^{j}(2)-S_{i}^{j}(1)],\dots,\frac{1}{S\cdot N}\sum_{i=1}^{S}\sum_{j=1}^{N}[S_{i}^{j}(AverTime)-S_{i}^{j}(1)]\right].
\end{eqnarray}
The Gaussian smoothed $Aver\mathcal{S}$ for each task is shown in Fig. \ref{Pupil_Size_all}.

Combining Fig. \ref{Pupil_Size_all} and the Analysis of Variance (ANOVA), we can obtain that the change in pupil size in Task 2 is significantly different from that in Task 1 $(F=344.37, p<0.01)$, that is, the IQA task can significantly affect the change in pupil size of subjects.
We also find that the pupils of subjects constrict first and then gradually dilate and recover whether they were observing images freely or evaluating image quality, and the times for the pupil to change from the constriction phase to the recovery phase in the two tasks are close.

\begin{figure*}[t]
\begin{minipage}[(a) ]{0.3\linewidth}
  \centering
  \centerline{\includegraphics[scale=0.32, trim=95 260 80 282, clip]{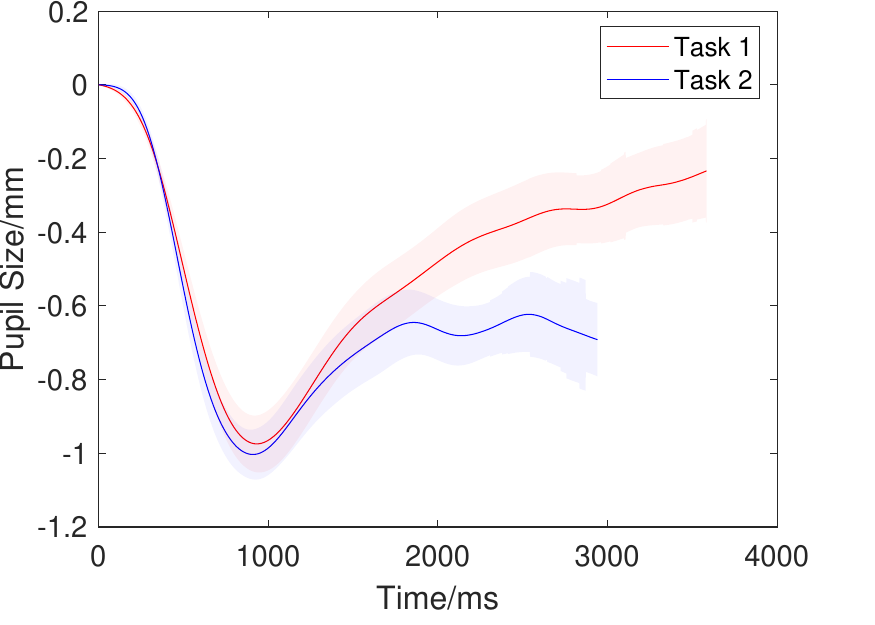}}
  \centerline{(a) JP2K}\medskip
  \end{minipage}
\begin{minipage}[(a) JP2K]{0.3\linewidth}
  \centering
  \centerline{\includegraphics[scale=0.32, trim=80 260 80 282, clip]{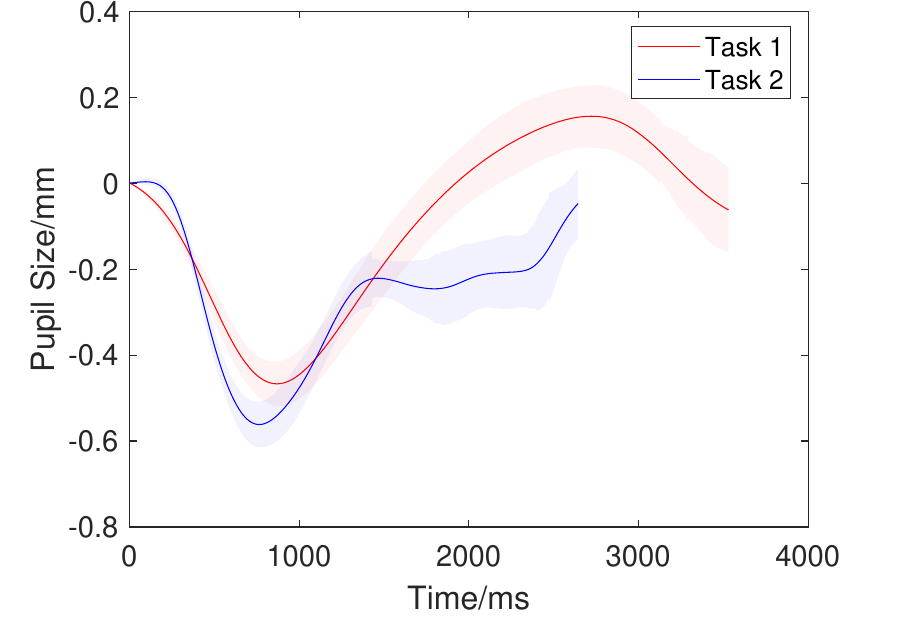}}
  \centerline{(b) JPEG}\medskip
\end{minipage}
\begin{minipage}[(a) ]{0.3\linewidth}
  \centering
  \centerline{\includegraphics[scale=0.32, trim=80 260 95 282, clip]{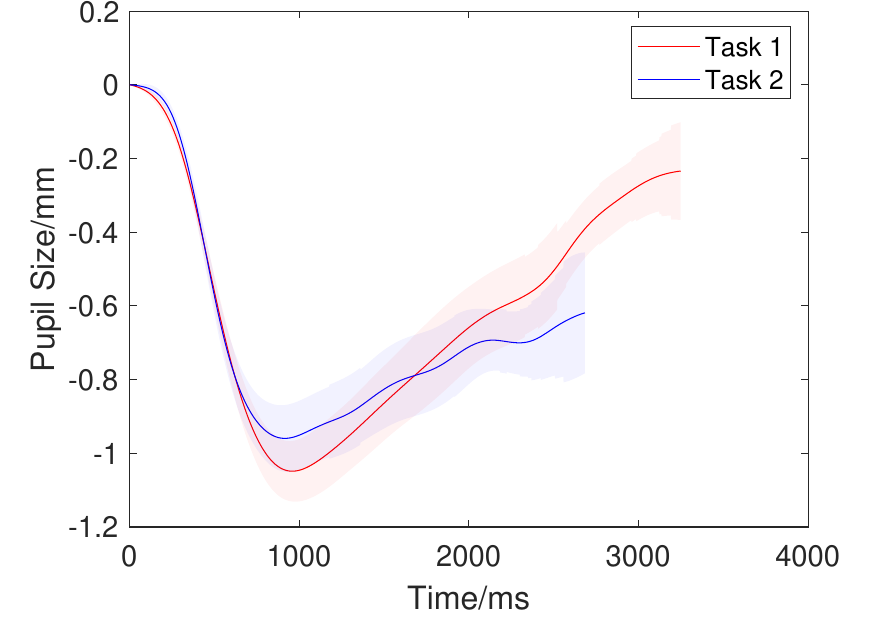}}
  \centerline{(c) WN}\medskip
\end{minipage}

\begin{minipage}[(a) ]{0.3\linewidth}
  \centering
  \centerline{\includegraphics[scale=0.32, trim=80 260 70 282, clip]{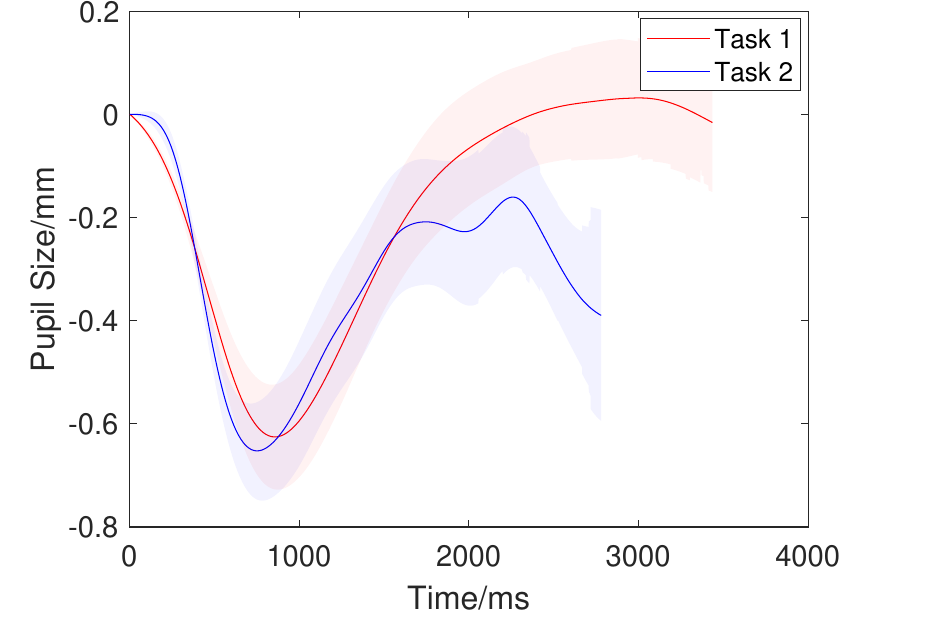}}
  \centerline{(d) GBlur}\medskip
\end{minipage}
\begin{minipage}[(a)]{0.3\linewidth}
  \centering
  \centerline{\includegraphics[scale=0.32, trim=80 260 80 282, clip]{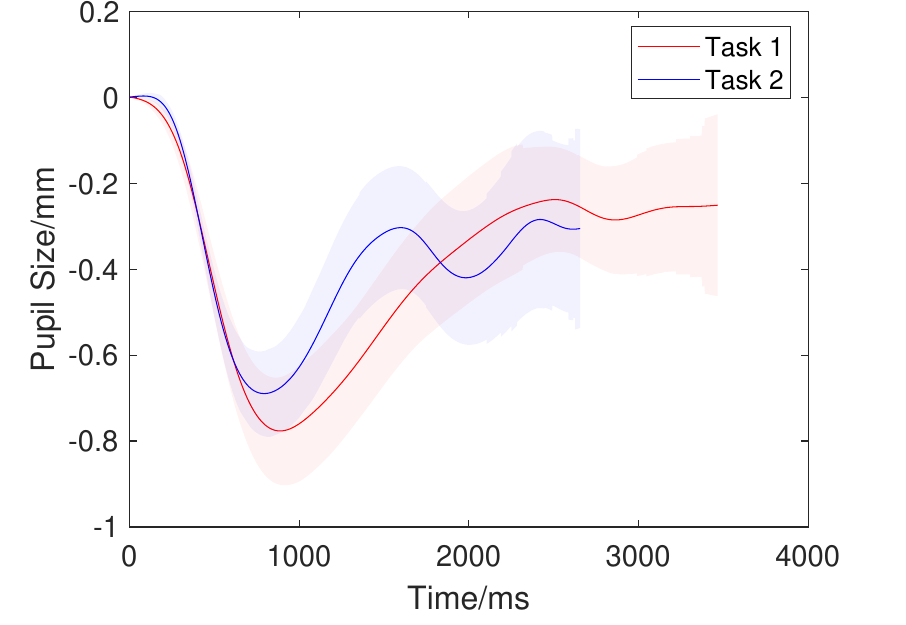}}
  \centerline{(e) FF}\medskip
\end{minipage}

\vspace{-10pt}
\caption{Changes in pupil size when subjects observed images with different distortion types in Task 1 and Task 2. Light colored areas represent standard errors.
}
\vspace{-5pt}
  \label{dist}
\end{figure*}

Similarly, we also analyze the changes in pupil size when subjects observed images with different distortion types in Task 1 and Task 2. The results are shown in Fig. \ref{dist}. 
Combining Fig. \ref{dist} and the ANOVA, we can obtain the following observations: the change in pupil size in Task 2 is significantly different from that in Task 1 when subjects observed images distorted by JP2K $(F=651.32, p<0.01)$; 
the change in pupil size in Task 2 is significantly different from that in Task 1 when subjects observed images distorted by JPEG $(F=1255.55, p<0.01)$;
the change in pupil size in Task 2 is significantly different from that in Task 1 when subjects observed images distorted by WN $(F=99.76, p<0.01)$;
the change in pupil size in Task 2 is significantly different from that in Task 1 when subjects observed images distorted by GBlur $(F=583.81, p<0.01)$; however, 
the change in pupil size in Task 2 is not significantly different from that in Task 1 when subjects observed images distorted by FF $(F=3.84, p=0.05)$. 
In addition to images distorted by FF, the IQA task can significantly affect the change in pupil size of subjects when they observed images distorted by JP2K, JPEG, WN, and GBlur.

To further study how the IQA task affects the change in pupil size, we divide the change in pupil size for each task into two phases: the initial constriction phase and the recovery phase after the constriction.

\subsection{Constriction Phase}
\begin{figure}[t]
\begin{minipage}[b]{1.0\linewidth}
\vspace{-20pt}
  \centering
  \vspace{1.5cm}
  \centerline{\includegraphics[scale=0.5, trim=80 250 80 268, clip]{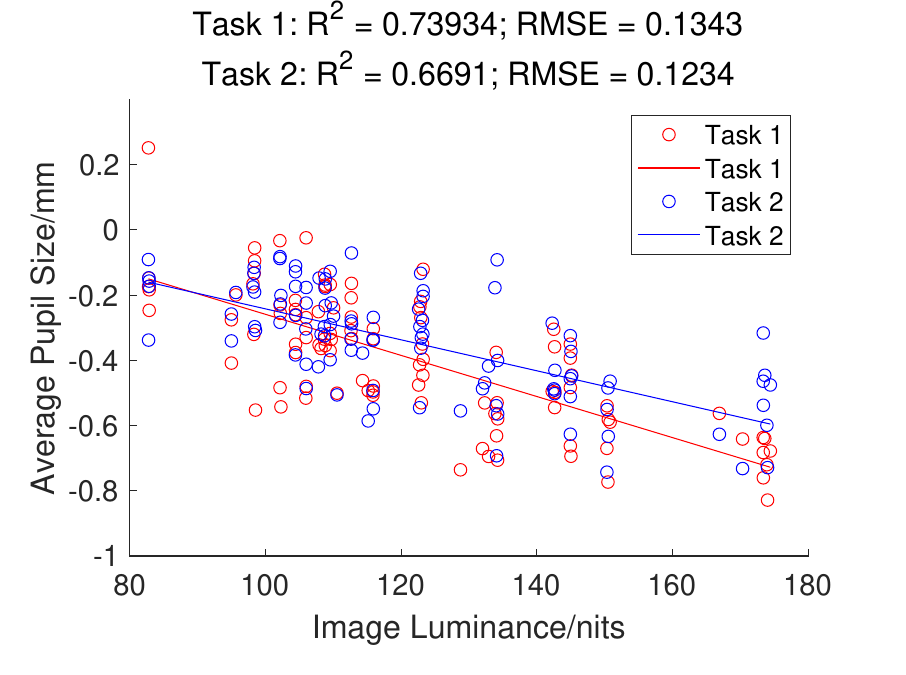}}
\end{minipage}
\vspace{-20pt}
\caption{Relationship between image luminance and the average pupil size at the pupil constriction phase.
Each circle represents an image, and the straight line is the result of linear regression.
}
\label{Image Luminance}
\end{figure}

From the perspective of the neural mechanism of pupil constriction, pupil constriction is mainly affected by light stimulation, and its neural channel is relatively simple, that is, when light is projected to the retina, nerve impulses are generated, and information is transmitted through the Edinger-Westphal (E-W) nucleus, which makes the pupil size decrease \cite{aston2005integrative}. 
Therefore, we assume that the pupil constriction in the two tasks is mainly caused by the change in screen luminance, that is, the sudden display of the image causes the pupil constriction.
Fig. \ref{Image Luminance} shows the relationship between the image luminance and the average pupil size at the pupil constriction phase for all images.
Each circle represents an image, and the straight line is the result of linear regression.
The $R^{2}$ and the root mean square error (RMSE) are used to test the goodness-of-fit of the regression results.
We have two observations. First, there is no significant difference in the average pupil size between the pupil constriction phases of the two tasks. Second, with the increase of image luminance, the degree of pupil constriction becomes obvious.
\subsection{Recovery Phase}
From Fig. \ref{Pupil_Size_all}, we can see that there is a significant difference in pupil size between the pupil recovery phases of the two tasks $(F=910.31, p<0.01)$. 
To further analyze this, we calculate the pupil recovery range $\Delta E$ of the recovery phase for each task:
\begin{eqnarray}\label{QuantileLoss}
{\Delta E}=\max\{Aver\mathcal{S}_r\}-\min\{Aver\mathcal{S}_r\},
\end{eqnarray}
where $Aver\mathcal{S}_r$ represents the average relative pupil size vector in the pupil recovery phase.
$\Delta E$ can be used to measure the degree of pupil recovery. The greater the value of $\Delta E$, the greater the degree of pupil recovery.
The pupil recovery ranges $\Delta E$ of Task 1 and Task 2 are 0.650 and 0.386, respectively.
Obviously, observing an image with the IQA task can significantly affect the degree of pupil recovery.


Next, we analyze what affects the degree of pupil recovery of subjects in the IQA task.
It is easy to find that the pupil oscillates between dilation and constriction in Task 2, which is caused by the visual attention allocation mechanism \cite{naber2013tracking}. This mechanism plays an important role in human visual perception: pupil dilation can increase the amount of visual information received by the brain, which indicates that people focus on global information;
pupil constriction can increase visual acuity, improve the ability to distinguish fine details, and selectively filter visual information when it enters the eyes. At this time, people are more concerned about local information. 
This shows that compared with observing freely, people are more likely to use the visual attention mechanism when evaluating image quality, which could help them analyze the details of the obtained image information.

Therefore, we propose that the IQA task affects the cognitive processes of people by activating the human visual attention mechanism.
It emphasizes that the visual attention mechanism is worth considering when designing objective IQA methods.

\begin{figure}[t]
\begin{minipage}[b]{1.0\linewidth}
\vspace{-40pt}
  \centering
  \vspace{1.5cm}
  \centerline{\includegraphics[scale=0.5, trim=80 250 80 280, clip]{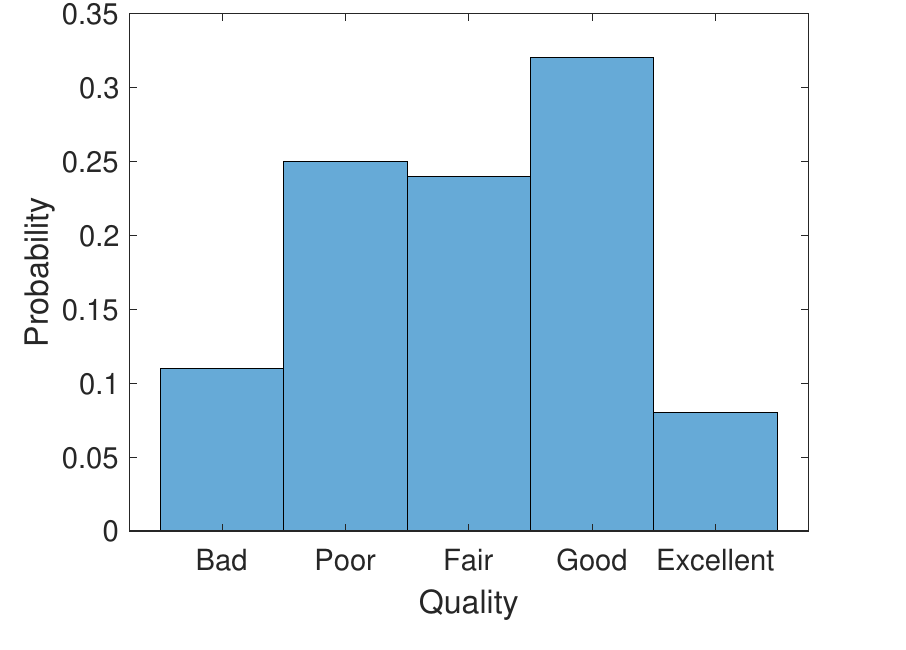}}
\end{minipage}
\vspace{-28pt}
\caption{Distribution of subjective quality of all images.}
\label{hist}
\end{figure}

\begin{figure}[t]
\begin{minipage}[b]{1.0\linewidth}
\vspace{-40pt}
  \centering
  \vspace{1.5cm}
  \centerline{\includegraphics[scale=0.5, trim=80 250 80 280, clip]{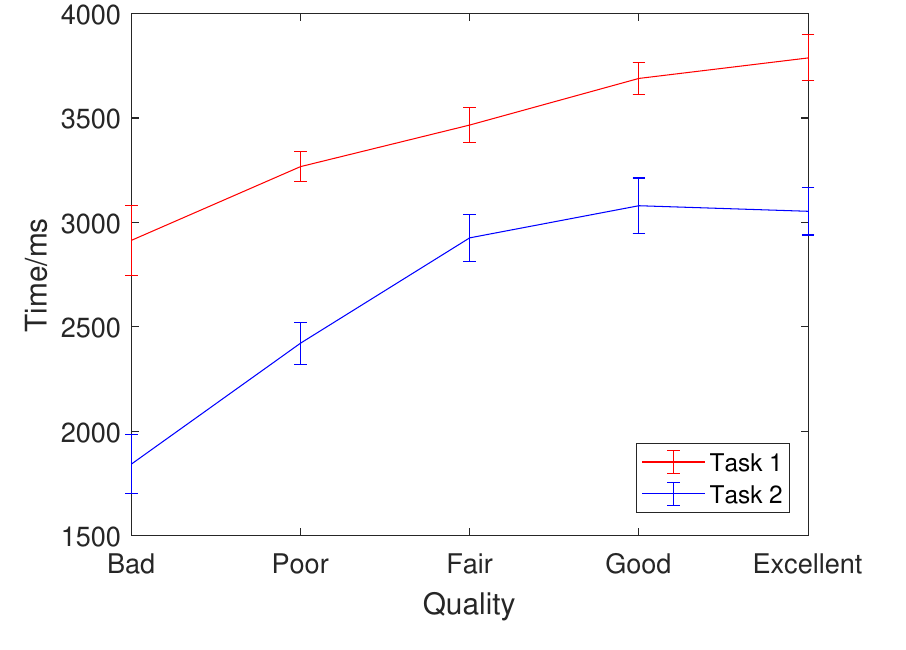}}
\end{minipage}
\vspace{-28pt}
\caption{Average observation time across all subjects for different image quality levels in Task 1 and Task 2.}
\label{Time}
\end{figure}

\section{Relationship with Image Quality}
\label{S4}
In this part, we specifically analyze the relationship between pupil data and image quality. Fig. \ref{hist}. shows the distribution of subjective quality of all images given by subjects in Task 2. 
First, we calculate the average observation time for different image quality levels in Task 1 and Task 2. The results are shown in Fig. \ref{Time}.
We have two observations from the figure. The first is that subjects would like to spend more time observing images freely than evaluating image quality, which is consistent with our findings in Section \ref{Analysis of Pupil Data}. The second is that the observation time of the image with bad quality is the shortest in the two tasks. This shows that when observing images freely, subjects were reluctant to spend too much time observing the image with bad quality. When evaluating image quality, subjects only needed a short observation time to give the quality score of the image with bad quality.

\begin{figure}[t]
\begin{minipage}[b]{1.0\linewidth}
\vspace{-40pt}
  \centering
  \vspace{1.5cm}
  \centerline{\includegraphics[scale=0.5, trim=80 250 80 280, clip]{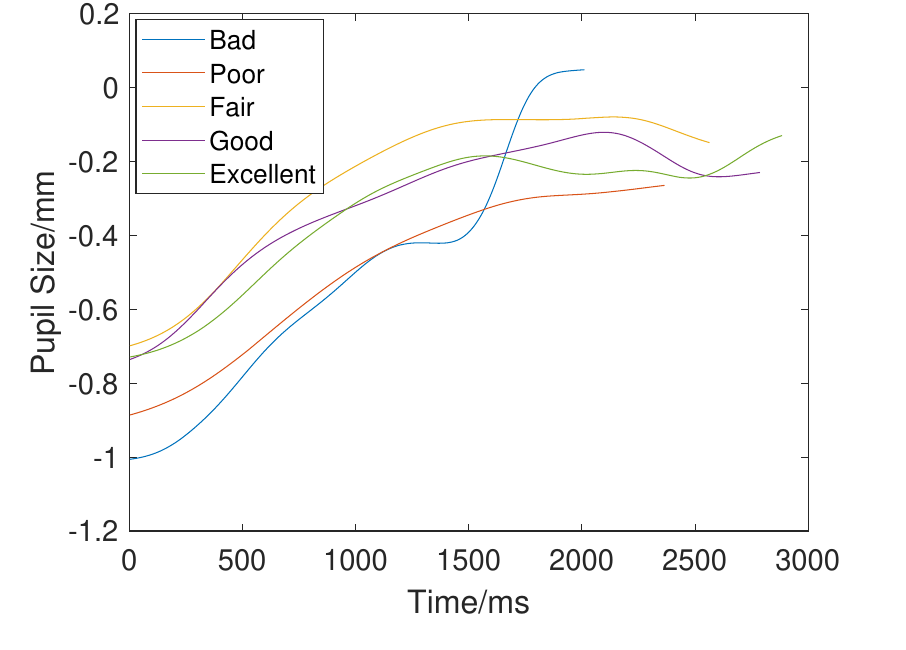}}
\end{minipage}
\vspace{-28pt}
\caption{Changes in pupil size for different image quality levels in the pupil recovery phase of Task 1.}
\label{1}
\end{figure}

\begin{figure}[t]
\begin{minipage}[b]{1.0\linewidth}
\vspace{-40pt}
  \centering
  \vspace{1.5cm}
  \centerline{\includegraphics[scale=0.5, trim=80 250 80 280, clip]{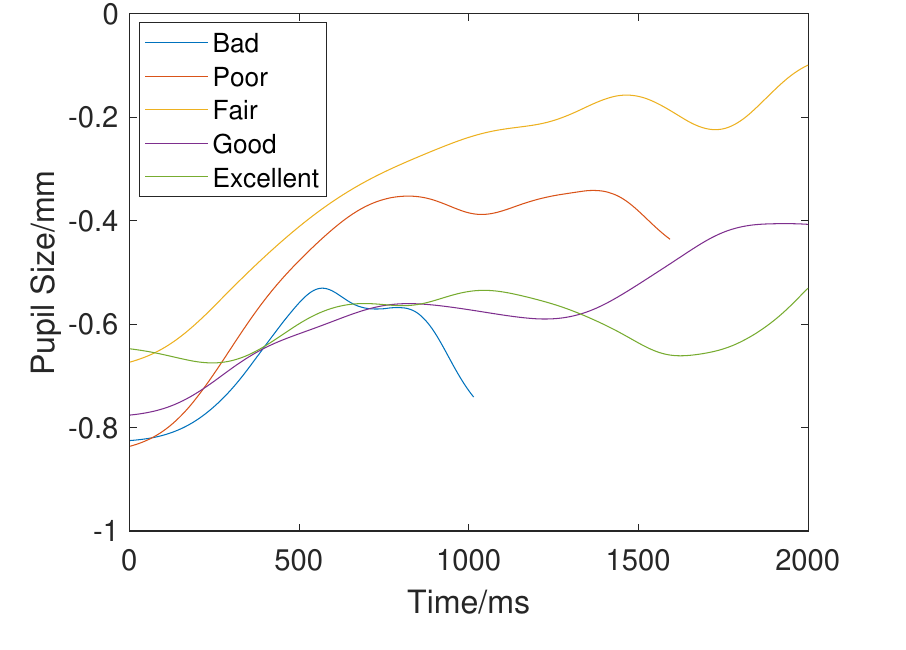}}
\end{minipage}
\vspace{-28pt}
\caption{Changes in pupil size for different image quality levels in the pupil recovery phase of Task 2.}
\label{2}
\end{figure}
According to the analyses in Section \ref{Analysis of Pupil Data}, we know that changes in pupil size of the two tasks can be divided into two phases. The first phase is closely related to image luminance, and the second phase is closely related to the task. Therefore, we mainly analyze the relationship between pupil size and image quality in the pupil recovery phase.

We analyze changes in pupil size for different image quality levels in the pupil recovery phase of Task 1 and Task 2. The results are shown in Fig. \ref{1} and Fig. \ref{2}. Comparing changes in pupil size between the two tasks, we can find that the oscillation between pupil dilation and constriction in Task 2 is more obvious than that in Task 1 for the same image quality level.
We further calculate the pupil recovery ranges $\Delta E$ of different image quality levels in the two tasks, and the results are shown in Table \ref{t1}. As can be seen from the table, the pupil recovery range of Task 2 is smaller than that of Task 1 for the same image quality level. This is consistent with the conclusion we have got in Section \ref{Analysis of Pupil Data}, that is, the IQA task can affect the cognitive processes of people.
\begin{table}[t]
\begin{center}
\caption{Pupil recovery ranges $\Delta E$ for different image quality levels in Task 1 and Task 2.
}
\label{t1}
\begin{tabular}{c||c|c|c|c|c}
  \hline\hline
$\Delta E$  (mm)& Bad&Poor&Fair&Good&Excellent \\
  \hline\hline
Task 1 &1.053&0.621&0.619&0.614&0.599\\\hline
Task 2&0.294&0.494&0.574&0.369&0.144\\
\hline\hline
\end{tabular}
\end{center}
\vspace{-4pt}
\end{table}

\begin{figure}[t]
\begin{minipage}[b]{1.0\linewidth}
\vspace{-40pt}
  \centering
  \vspace{1.5cm}
  \centerline{\includegraphics[scale=0.5, trim=80 250 80 270, clip]{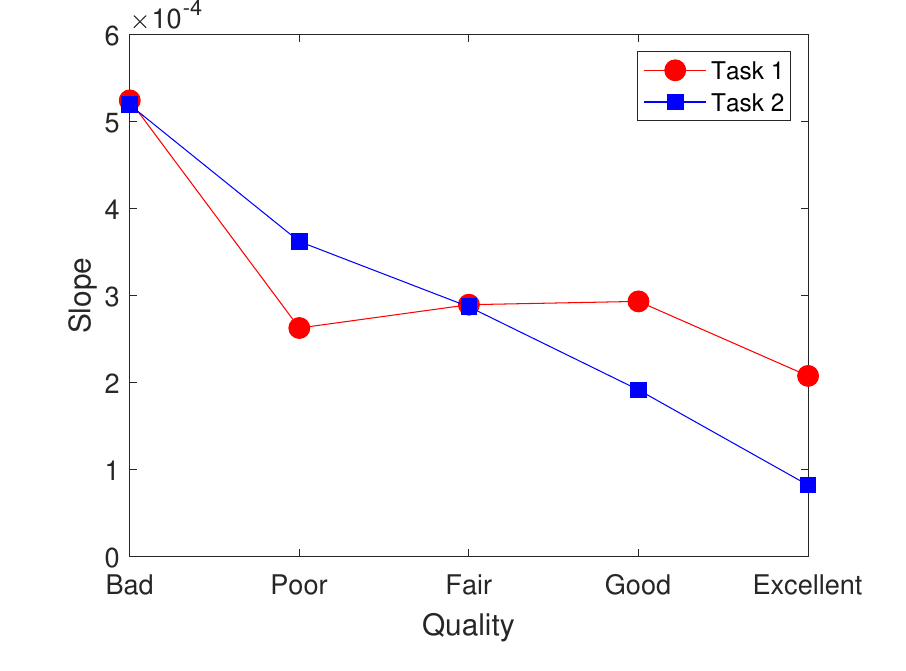}}
\end{minipage}
\vspace{-28pt}
\caption{Pupil recovery slopes for different image quality levels in Task 1 and Task 2.}
\label{slope}
\vspace{-4pt}
\end{figure}

In addition to the pupil recovery range, we also analyze the slopes of pupil recovery $K$ for different image quality levels in the pupil recovery phase of the two tasks, which can be defined as:
\begin{eqnarray}\label{QuantileLoss}
{K}=\Delta E/(T_{max}-T_{min}),
\end{eqnarray}
where
$T_{max}$ represents the time when the pupil size reaches the maximum in the pupil recovery phase, and $T_{min}$ represents the time when the pupil size reaches the minimum in the pupil recovery phase. 
$K$ can be used to measure the speed of pupil recovery. The higher the value of $K$, the faster the pupil recovers. Fig. \ref{slope} shows the pupil recovery slopes for different image quality levels in Task 1 and Task 2. We can obtain an interesting observation from the figure: in the IQA task, the quality of the observed image is closely related to the speed of pupil recovery. With the increase in image quality, the speed of pupil recovery decreases.

\section{Conclusion}
\label{S5}
This paper studies how the IQA task affects the cognitive processes of people from the perspective of pupil size, and the relationship between pupil size and image quality. Specifically, we first carry out a subjective experiment. In this experiment, we provide subjects with two tasks: free observation and IQA. In the free observation task, subjects do not need to perform any action. In the IQA task, subjects are asked to 
score images according to their overall impression of image quality. By analyzing the difference in pupil size between the two tasks, we find that people can activate the visual attention mechanism when evaluating image quality. We then analyze the relationship between the change in pupil size and image quality in the IQA task. We find that with the increase in image quality, the recovery speed after pupil contraction decreases. 
In the future, the research results of this paper can promote the development of objective IQA methods and improve subjective IQA methods.


\bibliographystyle{ACM-Reference-Format}
\bibliography{sample-base}

\appendix









\end{document}